\documentclass[12pt]{article}
\usepackage{amssymb,amsmath,amsthm}
\usepackage{graphicx}
\usepackage{natbib}
\usepackage[english]{babel}
\usepackage{booktabs}
\usepackage{enumitem}
\usepackage{xcolor}
\usepackage{tabularx}
\usepackage{url}
\usepackage{caption,setspace}
\newcommand{\blind}{0}

\newcommand{\by}{\boldsymbol{y}}
\newcommand{\bY}{\boldsymbol{Y}}
\newcommand{\bu}{\boldsymbol{u}}
\newcommand{\bU}{\boldsymbol{U}}
\newcommand{\btheta}{\boldsymbol{\theta}}
\newcommand{\bA}{\boldsymbol{A}}
\newcommand{\bB}{\boldsymbol{B}}
\newcommand{\bC}{\boldsymbol{C}}
\newcommand{\bD}{\boldsymbol{D}}
\newcommand{\bF}{\boldsymbol{F}}
\newcommand{\bI}{\boldsymbol{I}}
\newcommand{\bs}{\boldsymbol{s}}

\newcommand{\bzero}{\boldsymbol{0}}
\newcommand{\bP}{\boldsymbol{P}}
\newcommand{\bL}{\boldsymbol{L}}
\newcommand{\diff}{\mathop{}\!\mathrm{d}}

\title{\bf EM-Based Smooth Graphon Estimation \\ Using Bayesian and Spline-Based Approaches}
\if1\if0\blind 1\else\if2\blind 1\else0\fi\fi
{
	\author{Benjamin Sischka and G\"{o}ran Kauermann}
} \fi

\if1\blind
{
	\author{}
} \fi

\begin{document}

\def\spacingset#1{\renewcommand{\baselinestretch}
{#1}\small\normalsize} \spacingset{1}

\if0\blind
{
  \maketitle
  \let\thefootnote\relax\footnote{Benjamin Sischka is Research Assistant, Department of Statistics, Ludwig-Maximilians-Universit\"{a}t M\"{u}nchen, 80539 M\"{u}nchen, Germany (E-mail: \url{benjamin.sischka@stat.uni-muenchen.de}). G\"{o}ran Kauermann is Professor, Department of Statistics, Ludwig-Maximilians-Universit\"{a}t M\"{u}nchen, 80539 M\"{u}nchen, Germany (E-mail: \url{goeran.kauermann@stat.uni-muenchen.de}). 
  }
} \fi

\if1\blind
{
	\maketitle
} \fi

\if2\blind
{
	\maketitle
	\let\thefootnote\relax\footnote{Benjamin Sischka is Research Assistant, Department of Statistics, Ludwig-Maximilians-Universit\"{a}t M\"{u}nchen, 80539 M\"{u}nchen, Germany (E-mail: \url{benjamin.sischka@stat.uni-muenchen.de}). G\"{o}ran Kauermann is Professor, Department of Statistics, Ludwig-Maximilians-Universit\"{a}t M\"{u}nchen, 80539 M\"{u}nchen, Germany (E-mail: \url{goeran.kauermann@stat.uni-muenchen.de}). 
		
		The project was partially supported by the European Cooperation in Science and Tech\-nology [COST Action CA15109 (COSTNET)]. 
		
		This research did not receive any specific grant from funding agencies in the public, commercial, or not-for-profit sectors.
		
		Declarations of interest: none.
	}
} \fi
	
\bigskip
\begin{abstract}
	\noindent This paper proposes the estimation of a smooth graphon model for network data analysis using principles of the EM algorithm. The approach considers both variability with respect to ordering the nodes of a network and smooth estimation of the graphon by nonparametric regression. To do so, (linear) B-splines are used, which allow for smooth estimation of the graphon, conditional on the node ordering. This provides the M-step. The true ordering of the nodes arising from the graphon model remains unobserved and 
	Bayesian ideas are employed to obtain posterior samples given the network data. This yields the E-step. Combining both steps gives an EM-based approach for smooth graphon estimation. Unlike common other methods, this procedure does \textit{not} require the restriction of a monotonic marginal function. The proposed graphon estimate allows to explore node-ordering strategies and therefore to compare the common degree-based node ranking with the ordering conditional on the network. Variability and uncertainty are taken into account 
	using MCMC techniques. Examples and simulation studies support the applicability of the approach. 
\end{abstract}


\newpage
\spacingset{1.5}
\section{Introduction}
\label{sec:intro}
The analysis of network data has achieved increasing interest in the last years. \cite{Golden:10}, \cite{Hunter:12}, \cite{Fien:12} and \cite{Salter:12}, respectively, published survey articles demonstrating the state-of-the-art in the field. We also refer to \cite{Kol:09}, \cite{KolC:14} and \cite{Lush:13} for monographs in the field of statistical network data analysis, see also \cite{Kola:17}. The statistical workhorse models for network data are Exponential Random Graph Models (ERGM), Stochastic Block Models (SBM) and Latent Space Models. ERGMs make use of an exponential family distribution to model the network's adjacency matrix as a random matrix. This model class was proposed by \cite{Frank:86} and is extensively discussed in \cite{Snij:06}. SBMs as well as the latent space models are both model classes which make use of latent quantities relating to the nodes, that are employed to explore the network structure. They were proposed in their nowadays form by \cite{Holland:83} and \cite{Hoff:02}, respectively.

A different modeling strategy, which is also based on latent quantities, results through comprehending the network adjacency matrix $\bY \in \{0,1\}^{N \times N}$ to be generated by a so called graphon. The graphon as data generating model comes into play by assuming that we draw $N$ random variables 
\begin{align} 
	U_1,\ldots, U_N \overset{\text{\textit{i.i.d.}}}{\sim} \mbox{ Uniform}[0,1] 
	\label{eq:us}
\end{align}
and, given $U_i$ and $U_j$, simulate the network entries $Y_{ij}$ conditionally independently through 
\begin{align}
	Y_{ij}|U_i=u_i, U_j=u_j \sim \mbox{ Binomial}(1, w(u_i, u_j)).
	\label{eq:grm}
\end{align}
The function $w(,)$ is thereby called a graphon (= graph function). In case of undirected networks we additionally require symmetry so that $Y_{ij} = Y_{ji}$ (and hence in principle we assume $w(u,v) = w(v,u)$). We remain with the commonly used convention and exclude so called self-loops, which means $Y_{ii} = 0$ for $i=1,\ldots,N$.
Apparently, the graphon function $w(,)$ is not unique with regard to the data generating process (\ref{eq:grm}). This is because for any measure-preserving bijection $\varphi: [0,1] \rightarrow [0,1]$ the permuted graphon $w(\varphi(u), \varphi(v))$ yields the same model. More generally, as has been stated by \cite{Diac:08}, two graphons $w(,)$ and $w'(,)$ represent the same generating model if and only if there exist two measure preserving mappings -- not necessarily bijections -- $\tilde{\varphi}$ and $\tilde{\varphi}': [0,1] \rightarrow [0,1]$ such that $w(\tilde{\varphi}(u), \tilde{\varphi}(v)) = w'(\tilde{\varphi}'(u), \tilde{\varphi}'(v))$ for almost all $(u,v) \in [0,1]^2$. Some papers therefore add a further attribute to achieve uniqueness, e.g.\ see \cite{Bick:09} or \cite{Yangetal:14}. The common setting to do so is to postulate that 
\begin{align}
	g(u) = \int w(u,v) \diff v
	\label{eq:g}
\end{align}
is strictly increasing in $u$, which leads to the so called canonical representation of the graphon $w(,)$. Note that the distribution of $g(U_i)$ with $U_i$ following (\ref{eq:us}) can be interpreted as (asymptotic) distribution of the degree proportion. However, the additional condition (\ref{eq:g}) incorporated to tackle the identifiability issue can be problematic, especially in combination with other postulations. For reconstruction reasons commonly some kind of smoothness is assumed in the sense that $w(,)$ satisfies (at least piecewise) some Lipschitz or H\"{o}lder conditions of continuity, e.g.\ see \cite{WolfeOl:14}, \cite{ChAir:14} or \citeauthor{Gao:15a}\ (2015a). But then, considering an arbitrary continuous graphon, the transformation into its canonical representation neither usually retains the continuity nor are there necessarily any guarantees that such a canonical representation exists. Taking only graphon functions into account which are continuous in their canonical representation would apparently be a strong restriction of the generality of graphon models. We therefore consider the more general model class which does not include the canonical constraint. Still, we also discuss estimation under the restriction (\ref{eq:g}).

Graphon estimation for modeling network data has recently found attention in the statistical literature. A general discussion is given in \cite{Orbanz:14} and \cite{graphon:20} recently launched an \texttt{R} package for graphon estimation. Graphons can be related to ERGMs, at least for simple statistics like two-star or triangles, as shown in \cite{DiacChat:13}. \cite{He:15} make use of this connection and propose to use asymptotic properties of graphons to derive estimates in high dimensional ERGMs. Moreover, graphon models can be seen as a generalized model class which also includes SBMs and latent space models (e.g.\ see \citeauthor{Bickel:11}, \citeyear{Bickel:11}). \cite{WolfeOl:13} and \cite{Yangetal:14} discuss non-parametric estimation of graphons including tests on the validity of prespecified graphon shapes, see also \cite{ChAir:14}, \cite{Airoldi:13} or \cite{WolfeOl:14}. One of the most propagated strategies in the graphon estimation literature is to approximate the graphon through a SBM, see \cite{Choi:14}, \cite{Choi:17} or \cite{Klopp:17}. \citeauthor{Gao:15a}\ (2015a) discuss optimal graphon estimation for the SBM approximation. SBMs are, however, by definition discontinuous models, meaning that the corresponding graphon function is discontinuous and hence not smooth. In this paper, in contrast, we focus on smooth and hence continuous graphon functions. For a general discussion on graphons we refer to \cite{Bor:08}, \cite{lovasz:12}, \cite{Diac:08} or \cite{Bick:09}. 

Estimating graphons can generally be reduced to probability matrix estimation, where the goal is to gain information about the specific pairwise probabilities $\mbox{P}(Y_{ij} = 1 | U_i=u_i, U_j=u_j) = w(u_i,u_j)$. From this perspective, the $U_i$ are not well-defined components. This reduction -- considering the graphon values only at unspecific positions -- allows to circumvent the identifiability issue from above. \cite{Chatter:15} provides convergence results for general matrix estimation using singular value decomposition, see also \citeauthor{Gao:15a}\ (2015a), \cite{Zhang:15} or \cite{Xu:17}. \cite{Gao:16} extends this towards partially observed matrices/networks, see also \citeauthor{Gao:15b}\ (2015b) for a Bayesian approach. The cited work, however, yields estimates of specific graphon values at unspecific positions and thus does not lead to a smooth graphon estimate on the domain $[0,1]^2$ which is focused in this work.

In this paper we propose to use penalized linear B-splines for graphon estimation. This borrows ideas suggested in \cite{Kauermann:13} for copula estimation, since B-splines easily allow to accommodate side constraints such as symmtery in form of $w(u,v) = w(v,u)$ or, if required, condition (\ref{eq:g}) for the resulting estimate. This, in contrast, is difficult to accommodate in histogram or kernel based estimation. Penalized estimation with B-splines has thereby a long standing tradition in smooth estimation, starting with \cite{Eilers:96} and \citeauthor{Ruppert:09} (\citeyear{Ruppert:03}, \citeyear{Ruppert:09}), see also \cite{Wood:17}. We extend the idea here to graphon estimation. However, before this smoothing approach can be applied, we also need to fill the lack of information about the $U_i$. On the other hand, conditional on the observed network data $\by = \{y_{ij} \}_{i,j=1,\ldots,N}$, a presumption about the positions of the $U_i$ can only be made in relation to $w(,)$. Since both the implicitly supposed parameters of the B-spline approximation of $w(,)$ and the latent $U_i$ are unknown, this is a typical task for an EM type algorithm. Moreover, MCMC techniques can be applied in this context to approximate numerically the complex posterior distribution of the $U_i$, which together yields a MCEM algorithm.

The paper is organized as follows. Section~\ref{sec:EmpGraph} displays the main ideas of pursuing an EM based algorithm for smooth graphon estimation. Section~\ref{sec:EM} describes the procedure in detail. Section~\ref{sec:Simu} and Section~\ref{sec:realWorld} showcase results for both simulations and real-world data examples. A discussion concludes the paper.

\section{Graphon Representation and EM Motivation}
\label{sec:EmpGraph}
We assume that the graphon $w: [0,1]^2 \rightarrow [0,1]$ is a smooth function, meaning that $w(,)$ is sufficiently differentiable in both arguments. We call $w(,)$ to have a canonical representation if $g(u) = \int w(u,v) \diff v$ is strictly increasing, see, among others, \cite{ChAir:14}. We assume further that $w(,)$ is symmetric and generates a network of size $N$ through the following process. For $N$ independent uniform variables
$$ U_i \overset{\text{\textit{i.i.d.}}}{\sim} \text{ Uniform}[0,1], \quad i=1,\ldots,N $$
we obtain the symmetric network through
\begin{align}
	\mbox{P}(Y_{ij} = 1| U_i = u_i, U_j = u_j) = w(u_i, u_j)
	\label{eq:mod1}
\end{align}
for $1 \leq i < j \leq N$, where $Y_{ji} = Y_{ij}$ and $Y_{ii}=0$. The variables $U_i$ remain unobservable and as data we only obtain the observed network $\by$. This implies in particular that no information on the values of $U_i$ is given and hence estimation of $w(,)$ occurs to be difficult. We propose to tackle the estimation problem using an EM algorithm. In that sense, we calculate (or rather approximate by simulating) $\mbox{E}(\bU| \by)$ from which we derive an appropriate ordering, giving the E-step, which in turn allows to estimate $w(,)$ using smoothing techniques, providing the M-step. For the E-step we take a Bayesian view by looking at the posterior density of $\bU = (U_1,\ldots,U_N)$ given $\by$. Note that since $U_1,\ldots,U_N$ are $\text{\textit{i.i.d.}}$ uniform we have
\begin{align*}
	f(\bu|\by) \propto \prod_{\substack{i,j \\ j > i}} w(u_i, u_j)^{y_{ij}} (1-w(u_i,u_j))^{1-y_{ij}}.
\end{align*}
If we look at the univariate distribution of a single variable $U_k$ given the entire network $\by$, this results through
\begin{align}
	f_k (u_k|\by) \propto \int \ldots \int \underset{j > i}{\prod_{i,j}} w(u_i, u_j)^{y_{ij}} (1-w(u_i, u_j))^{1-y_{ij}} \diff u_1 \ldots \diff u_{k-1} \diff u_{k+1} \ldots \diff u_N.
	\label{eq:int1}
\end{align}
Apparently, both the joint and the marginal posterior distribution are too complex to be calculated analytically, in particular if $N$ is large. We will therefore explore (\ref{eq:int1}) by pursuing a Bayesian approach using MCMC techniques. Details are given in the next sections. Letting $\hat{u}_k^{(m)}$ denote the rescaled ranks of the posterior means (see equation (\ref{eq:u_em}) below) of the MCMC samples in the $m$th step of the EM algorithm, we then maximize the likelihood resulting from (\ref{eq:mod1}) with $u_i$ replaced by $\hat{u}_i^{(m)}$. To do so, we make use of penalized B-spline smoothing, meaning that we replace the unknown graphon $w(,)$ by an approximate spline base representation $\bB(,) \btheta$, where $\bB(,)$ is a bivariate spline and $\btheta$ is the vector of unknown spline coefficients to be estimated in the M-step. Both steps the E- and the M-step will be introduced in detail below. Before doing so, however, we propose a simple first E-step, which also serves as initial step of the corresponding EM iteration. 

Even though the probability model (\ref{eq:mod1}) used for the construction of networks is simple, it can not directly be used for estimation. The reason is that variables $U_i$ are unobservable and hence can not directly be employed to estimate the graphon $w(,)$. Instead, in the recent literature the graphon $w(,)$ is usually estimated by smoothing the observed adjacency matrix $\by$. The preceding rearrangement of the nodes thereby should effect an appropriate reordering of the rows and columns which in turn should reflect the ordering of the $U_i$. Let $\psi: \{1,\ldots,N\} \rightarrow \{1,\ldots,N\}$ be a permutation such that
\begin{align*}
U_{\psi (i)} \leq U_{\psi (j)}
\end{align*} 
with $i < j$. This means $U_{\psi(i)}= U_{(i)}$, where $U_{(1)} \leq U_{(2)} \leq \ldots \leq U_{(N)}$ define the ordered variables $U_i$. Note that since $U_i$, $i=1,\ldots,N$ are not observable, we can also not observe $\psi()$ which therefore needs to be estimated. A common approach is to make use of the degree. Let therefore $\hat{\psi} : \{1,\ldots,N\} \rightarrow \{1,\ldots,N\}$ be a permutation such that 
\begin{align}
	\mbox{\textit{degree}} ( \hat{\psi} ( i )) \leq \mbox{\textit{degree}} ( \hat{\psi}( j ))	
	\label{eq:truePerm}
\end{align}
for $i < j$. Note that $\hat{\psi}()$ can serve as an initial estimate for $\psi()$, although it requires the assumption that $g()$ from (\ref{eq:g}) is strictly increasing. We therefore use it as initial E-step and define the corresponding resulting initial prediction for $U_j$ based on this simple sorting through 
\begin{align}
	\hat{u}_j^{(0)} = \frac{\mbox{\textit{rank}}(\mbox{\textit{degree}}(j))}{N+1},
	\label{eq:uinit}
\end{align}
where $\mbox{\textit{rank}}(\mbox{\textit{degree}}(j))$ is the rank from smallest to largest of the $j$th element of the tuple $(\mbox{\textit{degree}}(i):\; \allowbreak i =1 ,\ldots, N)$. This is equivalent to define $\hat{u}_{\hat{\psi}(j)}^{(0)} = j/(N+1)$, where $i/(N+1)$, $i=1,\ldots,N$ represent the expected values of $N$ ordered independently $\text{Uniform}[0,1]$ distributed variables. \cite{ChAir:14} prove -- in case of graphons with canonical representation -- asymptotic convergence rates for $\hat{\psi}()$ towards $\psi()$, meaning that $|\psi(j) - \hat{\psi}(j)|/N \rightarrow 0$ for all $j = 1,\ldots, N$.

We can now replace $w(,)$ in (\ref{eq:mod1}) by its empirical version $\hat{w}^{(0)}(,)$ which is defined through
\begin{align*}
	\hat{w}^{(0)}(u,v)=y_{\hat{\psi}(\lceil u N \rceil) \hat{\psi}(\lceil v N \rceil)},
\end{align*}
where $\lceil u N \rceil$ defines the smallest value which is greater or equal to $u N$. Note that $\hat{w}^{(0)}(,)$ just mimics the ordered adjacency matrix scaled towards the unit square. These calculations provide the initial steps in the subsequently introduced EM algorithm.

\section{EM Algorithm for smooth Graphons}
\label{sec:EM}

\subsection{Bayesian Approach for the E-Step}
\label{subsec:BayApp}

We pursue the E-step by exploiting the posterior distribution of $\bU$. This is done by constructing an appropriate MCMC Gibbs sampling scheme based on the full posterior distribution of $U_k$. Note that by conditioning on $\by$ and all $U_j$ except of $U_k$ one gets 
\begin{align}
	f_k(u_k|u_1, \ldots, u_{k-1}, u_{k+1}, \ldots, u_N, \by) \propto \prod_{j \neq k} w(u_k, u_j)^{y_{kj}} (1-w(u_k, u_j))^{1-y_{kj}}. 
	\label{eq:Gibb1}
\end{align}
We pretend in this section that the graphon $w(,)$ is known, which is the general setting in the E-step. This allows to easily sample from (\ref{eq:Gibb1}) using Gibbs sampling as MCMC technique. To do so, we assume $\bu^{<t>} = (u^{<t>}_{1},\ldots,u^{<t>}_{N})$ to be the current state of the Markov chain. To update the $k$th component we then set $u^{<t+1>}_{j} := u^{<t>}_{j}$ for $j \neq k$, while component $u_k$ is updated by drawing from (\ref{eq:Gibb1}). To pursue this, we make use of a normal proposal using a logit link. To be specific, let $z^{<t>}_{k} = \log(u^{<t>}_{k}/(1-u^{<t>}_{k})) = \text{ logit }(u^{<t>}_{k})$. We then propose to draw $z^*_k = z^{<t>}_{k} + \mbox{Normal}(0, \sigma^2)$ and set $u_k^* = \text{ logit}^{-1} (z_k^*) = \exp(z^*_k) \big/ (1+ \exp(z_k^*))$. 
Hence, the proposal density for $U_k$ is proportional to 
\begin{align*}
	q(u_k^* | u^{<t>}_{k}) &= \frac{\partial u^*_k}{\partial z^*_k} \phi(z^*_k|z^{<t>}_{k}) \\
	&\propto \frac{1}{u^*_k (1-u_k^*)} \exp \left(-\frac{1}{2} \frac{(\text{logit } (u_k^*) - \text{ logit }(u^{<t>}_{k}))^2}{\sigma^2}\right),
\end{align*}
where $\phi()$ is the standard normal density. Consequently, the ratio of proposals equals
\begin{align*}
	\frac{q_k(u^{<t>}_{k} | u^*_k)}{q_k(u_k^*|u^{<t>}_{k})} = \frac{u^*_k(1-u^*_k)}{u^{<t>}_{k} (1- u^{<t>}_{k})}.
\end{align*}
The proposed value $u^*_k$ is accepted (and hence we set $u^{<t+1>}_{k} := u_k^*$) with probability
\begin{align*}
	\min \left\{ 1, \quad \prod_{j \neq k} 
	\vphantom{\left(  \frac{(1-w(u^*_k, u^{<t>}_{j}))}{(1-w(u^{<t>}_{k}, u^{<t>}_{j}))}   \right)^{1-y_{kj}}} \right.  \left[ \left(  \frac{w(u_k^*, u^{<t>}_{j})}{w(u^{<t>}_{k}, u^{<t>}_{j})}  \right)^{y_{kj}}   
	\left(  \frac{1-w(u^*_k, u^{<t>}_{j})}{1-w(u^{<t>}_{k}, u^{<t>}_{j})}   \right)^{1-y_{kj}} \right] \left. \vphantom{\left(  \frac{(1-w(u^*_k, u^{<t>}_{j}))}{(1-w(u^{<t>}_{k}, u^{<t>}_{j}))}   \right)^{1-y_{kj}}}
	\frac{u^*_k(1-u^*_k)}{u^{<t>}_{k} (1- u^{<t>}_{k})}    \right\}.
\end{align*}
If we do not accept $u_k^*$, we set $u^{<t+1>}_{k} := u^{<t>}_{k}$. Based on the resulting Markov chain we can estimate the posterior mean $\mbox{E}(U_k | \by)$ by taking the mean of the simulated values, observing an appropriate burn-in phase of the algorithm. To be specific, we use the MCMC sequence to estimate the posterior mean in the $m$th step of the EM algorithm through
\begin{align*}
	\bar{u}_k^{(m)} = \frac{1}{n} \sum_{s=1}^n u^{<s \cdot N \cdot r>}_{k},
\end{align*}
where $r \in \mathbb{N}$ describes a thinning factor and $n$ is the number of MCMC states which are taken into account. We then use the ranks of the posterior means to reorder the network matrix accordingly. This corresponds to setting the missing values of $U_j$ according to (\ref{eq:uinit}) to
\begin{align}
	\hat{u}_j^{(m)} = \frac{\mbox{\textit{rank}} (\bar{u}_j^{(m)}) }{N+1}. 
	\label{eq:u_em}
\end{align}
The Gibbs sampling approach is straightforward and simple but requires the knowledge of the graphon $w(,)$. Apparently, in the EM algorithm we replace $w(,)$ in the formula above by the current estimate resulting through the M-step, which is described subsequently. We define the final estimate resulting through (\ref{eq:u_em}) as $ \hat{u}_j^{\text{\textit{EM}}}$.

\subsection{Spline based Graphon Estimation for the M-Step}
\label{subsec:splines}

\subsubsection{Linear B-Splines}
\begin{figure}
	\centering
	\includegraphics[scale = 0.55]{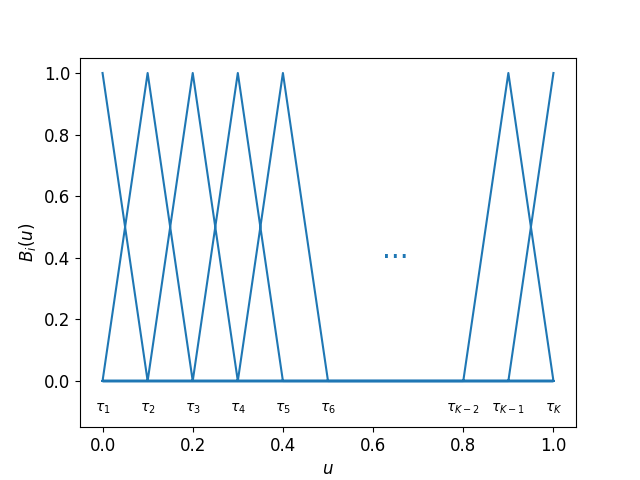}
	\caption{Normalized univariate linear B-spline basis for the approximation of the graphon. The (equidistant) inner knots are denoted by $\tau_j$ with $j=1,\ldots,K$.}
	\label{fig:splBas}
\end{figure}

For smooth estimation of the graphon $w(,)$ we first formulate a spline based approximation through 
\begin{align}
	w_{\btheta}^{\text{\textit{spline}}}(u,v) = \bB(u,v) \, \btheta = \left[ \bB(u) \otimes \bB(v) \right] \btheta,
	\label{eq:approx}
\end{align}
where $\bB(u) \in \mathbb{R}^{1 \times K}$ is a linear B-spline basis on $[0,1]$, normalized to have maximum value 1, see Figure~\ref{fig:splBas}. Parameter vector $\btheta \in \mathbb{R}^{K^2 \times 1}$ is indexed through
\[
\btheta = \left( \theta_{11},\ldots, \theta_{1K}, \theta_{21}, \ldots, \theta_{K1},\ldots,, \theta_{KK} \right)^\top.
\]
Using (\ref{eq:approx}) we obtain the likelihood
\[
l(\btheta) = \sum\limits_{i} \sum\limits_{j \neq i} \left[ y_{ij} \, \log \left( \bB_{ij} \btheta \right) + \left( 1- y_{ij} \right) \, \log \left( 1 - \bB_{ij} \btheta \right) \right],
\]
where $\bB_{ij} = \bB(u_i) \otimes \bB(u_j)$. Taking the derivative leads to the score function
\[
\bs ( \btheta) = \sum\limits_{i} \sum\limits_{j \neq i} \bB_{ij}^\top \left( \frac{y_{ij}}{w_{\btheta}^{\text{\textit{spline}}}(u_i,u_j)} - \frac{1-y_{ij}}{1-w_{\btheta}^{\text{\textit{spline}}}(u_i,u_j)} \right).
\]
Moreover, taking the expected second order derivative leads to the Fisher matrix
\[
\bF(\btheta) = \sum\limits_{i} \sum\limits_{j \neq i} \bB_{ij}^\top \bB_{ij} \left[ w_{\btheta}^{\text{\textit{spline}}} \left( u_i,u_j \right) \cdot \left( 1 - w_{\btheta}^{\text{\textit{spline}}} \left( u_i,u_j \right) \right) \right]^{-1}.
\]
Our intention is to maximize $l(\btheta)$, which could be simply done by Fisher scoring. The resulting maximizer does, however, not by default lead to a proper estimate, meaning to fulfill symmetry and boundedness. Furthermore, in case we aim to estimate a graphon with canonical representation, we need to incorporate the constraint from (\ref{eq:g}). We therefore impose additional (linear) side constraints on $\btheta$. Considering the identifiability issue we get the marginal function with (\ref{eq:approx}) through
\begin{align}
	g_{\btheta}^{\text{\textit{spline}}}(u) = \left[ \bB(u) \otimes \int\limits_0^1 \bB(v) \diff v \right] \btheta.
	\label{eq:G_1}
\end{align}
For normalized B-splines we can easily calculate the integral and for a standardized basis with equidistant knots we obtain
\begin{align*}
	\int\limits_0^1 \bB(v) \diff v &= \left( \int\limits_0^1 B_1(v) \diff v, \int\limits_0^1 B_2(v) \diff v, \ldots, \int\limits_0^1 B_K(v) \diff v \right) \\
	&= \underbrace{\frac{1}{K-1} \left( \frac{1}{2}, 1, \ldots, 1, \frac{1}{2} \right)}_{ =: \bA}.
\end{align*}
This allows to rewrite (\ref{eq:G_1}) to $g_{\btheta}^{\text{\textit{spline}}}(u) = \left[ \bB(u) \otimes \bA \right] \btheta$. Hence, the marginal function $g_{\btheta}^{\text{\textit{spline}}}()$ is also expressed as a linear B-spline and a monotonicity constraint is easily accommodated by postulating monotonicity at the knots $\tau_1,\ldots, \tau_K$. That is we need
\begin{align}
	g_{\btheta}^{\text{\textit{spline}}} \left( \tau_l \right) - g_{\btheta}^{\text{\textit{spline}}} \left( \tau_{l-1} \right) \geq 0 \quad \Leftrightarrow \quad \left[ \left( \bB \left( \tau_l \right) - \bB \left( \tau_{l-1} \right) \right) \otimes \bA \right] \btheta \geq 0
	\label{eq:canonical}
\end{align} 
for $l = 2,\ldots,K$, which is a simple linear constraint on the coefficient vector. 

Imposing symmetry on the graphon can also easily be accommodated as linear constraints $\theta_{pq} = \theta_{qp}$ for $p \neq q$. Finally, we need $w_{\btheta}^{\text{\textit{spline}}} (,)$ being bounded to $[0,1]$, which is again a simple linear constraint. All in all, we can write the side constraints as $\bC \btheta \geq \bzero$ and $\bD \btheta = \bzero$ for matrices $\bC$ and $\bD$ chosen accordingly, where the constraint from (\ref{eq:canonical}) for the canonical representation can be added if required. With the above linear constraints and the maximization task of $l(\btheta)$ we obtain an (iterated) quadratic programming problem, which can be solved using standard software (see e.g.\ \citeauthor{cvxopt}, \citeyear{cvxopt} or \citeauthor{quadprog}, \citeyear{quadprog}).

\subsubsection{Penalized Estimation}
Following ideas from the penalized spline estimation (see \citeauthor{Eilers:96}, \citeyear{Eilers:96} or \citeauthor{Ruppert:09}, \citeyear{Ruppert:09}) we may additionally impose a penalty on the coefficients to achieve smoothness. This is necessary since we intend to choose $K$ large and unpenalized estimation will lead to wiggled estimates. We refer to \cite{Eilers:96} for a motivation of penalized spline estimation. To do so, we penalize the difference between ``neighbouring'' elements of $\btheta$ to achieve smoothness. Let therefore
\[
\bL = \begin{pmatrix}
1 & -1 & \phantom{-}0 & \phantom{-}\ldots & \phantom{-}0 \\
0 & \phantom{-}1 & -1 & \phantom{-}\ldots & \phantom{-}0 \\
\vdots & \multicolumn{2}{c}{\ddots} & & \phantom{-}\vdots \\
0 & \phantom{-}\ldots & \phantom{-}0 & \phantom{-}1 & -1 \\
\end{pmatrix} \in \mathbb{R}^{(K-1) \times K}
\]
be the first order difference matrix. We then penalize $\left[ \bL \otimes \bI \right] \btheta$ and $\left[ \bI \otimes \bL \right] \btheta$, where $\bI$ is the identity matrix of appropriate size. This is leading to the penalized likelihood
\[
l_{\bP} (\btheta, \lambda) = l (\btheta) - \frac{1}{2} \lambda \btheta^\top \bP \btheta ,
\] 
where $\bP = \left( \bL \otimes \bI \right)^{\top} \left( \bL \otimes \bI \right) +  \left( \bI \otimes \bL \right)^{\top} \left( \bI \otimes \bL \right) $ and $\lambda$ serves as smoothing parameter. The corresponding penalization score function is given through 
\[
\bs_{\bP}(\btheta, \lambda) = \bs(\btheta) - \lambda \bP \btheta
\] 
and the penalized Fisher matrix in the form 
\[
\bF_{\bP} (\btheta,\lambda) = \bF (\btheta) + \lambda \bP.
\] 
The estimate apparently depends on the penalty parameter $\lambda$, which is suppressed in the notation. Setting $\lambda \rightarrow 0$ gives an unpenalized fit while setting $\lambda \rightarrow \infty$ leads to a constant graphon, i.e.\ an Erd\H{o}s-R\'{e}nyi model. The smoothing parameter $\lambda$ therefore needs to be chosen data driven. We here follow \cite{Kauermann:13} and make use of the Akaike Information Criterion (AIC) (\citeauthor{Hurvich:89}, \citeyear{Hurvich:89}, see also \citeauthor{Burnham:10}, \citeyear{Burnham:10}). To do so, we define the corrected AIC through
\[
AIC_c (\lambda) = - 2 \, l ( \hat{\btheta}_{\bP} ) + 2 \, df (\lambda ) + \frac{2 \, df( \lambda ) \left( df (\lambda) + 1\right) }{ \left( N (N-1) \right) - df( \lambda ) - 1} ,
\]
where $\hat{\btheta}_{\bP}$ is the penalized parameter estimate and $df(\lambda)$ is the degree of the model, which we define in the usual way as trace of the product of the inverse penalized Fisher matrix and the unpenalized Fisher matrix. To be specific
\[
df(\lambda ) = tr \left\{ \bF_{\bP}^{-1} ( \hat{\btheta}_{\bP}, \lambda ) \bF ( \hat{\btheta}_{\bP} ) \right\}.
\]
We subsequently denote with $\hat{w}^{(1)}(,)$ and $\hat{w}^{\text{\textit{EM}}}(,)$ the penalized B-spline estimates of $w_{\btheta}^{\text{\textit{spline}}}(,)$ in the first and the final EM step, meaning that $\hat{w}^{(1)}(,)$ and $\hat{w}^{\text{\textit{EM}}}(,)$ are based on $\hat{\bu}^{(0)}$ and $\hat{\bu}^{\text{\textit{EM}}}$, respectively. 

As it is well known, the EM algorithm can be trapped at a local maximum of the likelihood (or, in this case, at a local minimum of the corrected AIC). Thus, depending on the specific data situation, it might be recommendable to repeat the algorithm several times to achieve a better fit.

\subsection{Information on Ranking}
\label{subsec:infonrank}
Note that $U_i$ is uniform and it is helpful to order $U_i$ such that $U_{(1)} \leq U_{(2)} \leq \ldots \leq U_{(N)}$. Considering the degree based ordering $\hat{\psi}()$ from (\ref{eq:truePerm}) as representative, this allows to define the (degree related) ranking density $f_{\hat{\psi}(k)}(u_{\hat{\psi}(k)}|\by)$. For the sake of simplicity, we subsequently collapse $\hat{\psi}(k)$ to $(k)$ so that henceforth $f_{(k)}(u_{(k)}|\by)$ and $U_{(k)}$ refer to the node with the $k$th lowest degree. For other quantities the notation applies accordingly. This ranking density, however, again is difficult or even impossible to calculate analytically. Note that the full posterior density is given through
\begin{align}
	f^{}_{(k)}(u_{(k)} | \bu_{(-k)}, \by) \propto \prod_{j \neq k} w^{}(u_{(k)}, u_{(j)})^{y_{(k)(j)}} (1- w^{}(u_{(k)}, u_{(j)}))^{1-y_{(k)(j)}},
	\label{eq:Gibb1Sub}
\end{align}
where $y_{(k)(j)}$ is the network entry which refers to the nodes $k$ and $j$ after the permutation according to the degree based ordering. 
In that regard, the MCMC sequence (after appropriate thinning) provides information about the posterior distribution of $\bU$ given the network $\by$. Hence, for the marginal posterior distribution of $U_{(k)}$ we can follow a Monte Carlo integration approach, see e.g.\ \citeauthor{GelSmith:90} (\citeyear{GelSmith:90}, sec. 2.2 and 2.3), and calculate
\begin{align}
	f_{(k)}^{} (u_{(k)}| \by) \approx \frac{1}{n} \sum_{s=1}^n f_{(k)}^{} (u_{(k)}|\bu^{<s\cdot N \cdot r>}_{ (-k)}, \by),
	\label{eq:postDist}
\end{align}
where $\bu^{<t>}_{ (-k)} = (u^{<t>}_{(1)}, \ldots, u^{<t>}_{(k-1)}, u^{<t>}_{(k+1)}, \ldots, u^{<t>}_{(N)})$ is the $t$th state of the Gibbs sampling sequence without the $k$th component after degree related permutation, $r \in \mathbb{N}$ describes a thinning factor and $n$ is the number of MCMC states which are taken into account. For this purpose, the unspecified normalizing constant in (\ref{eq:Gibb1Sub}) can be approximated through a Riemann sum, since we assume $w^{}(,)$ to fulfill certain continuity properties. Considering $w^{}(,)$ as unknown again, we employ $\hat{w}^{(1)}(,)$ or $\hat{w}^{\text{\textit{EM}}}(,)$ as estimate in (\ref{eq:Gibb1Sub}) and obtain $\hat{f}_{(k)}^{(1)} (u_{(k)}| \by)$ or $\hat{f}_{(k)}^{\text{\textit{EM}}} (u_{(k)}| \by)$ as estimate of (\ref{eq:postDist}), respectively.

\section{Simulation Studies}
\label{sec:Simu}
\begin{figure}
	\centering
	\begin{minipage}[t]{1\textwidth}
		\centering
		\begin{minipage}[t]{0.48\textwidth}
			\centering
			\includegraphics[width=.99\textwidth]{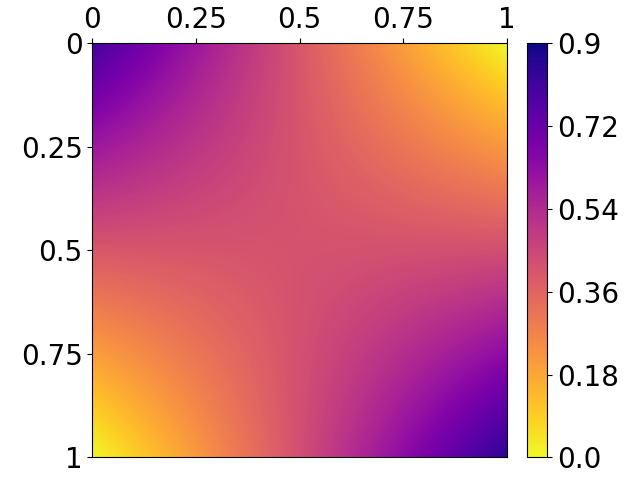}
		\end{minipage}
		\hspace*{0.1cm}
		\begin{minipage}[t]{0.48\textwidth}
			\centering
			\includegraphics[width=.99\textwidth]{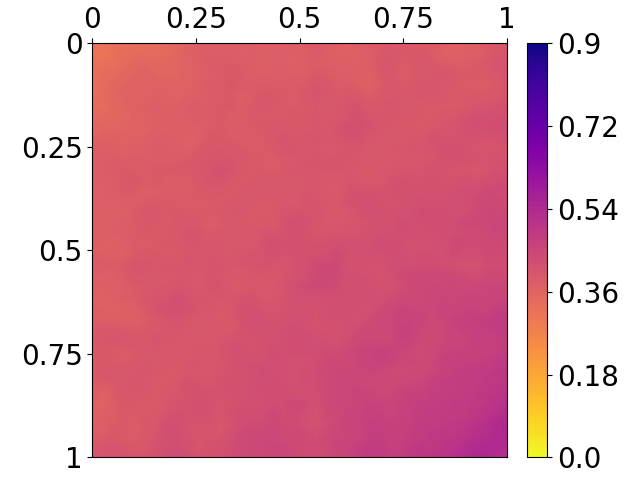}
		\end{minipage}
		\vspace*{0.3cm}
		
		\begin{minipage}[c]{0.49\textwidth}
			\centering
			\hspace*{-1.1cm}\includegraphics[width=.87\textwidth, height=.75\textwidth]{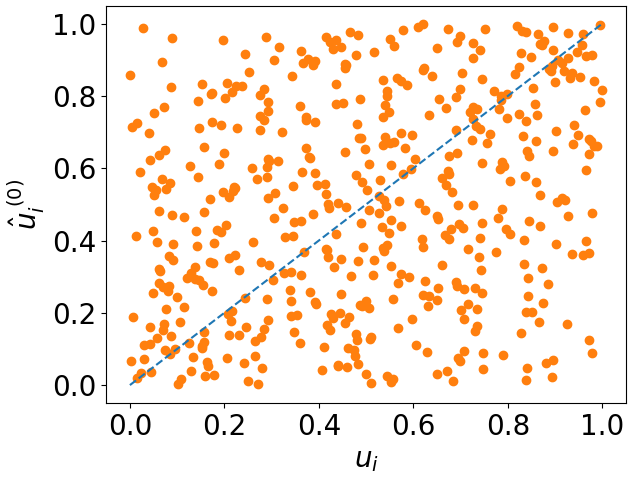}
		\end{minipage}
		\begin{minipage}[c]{0.49\textwidth}
			\centering
			\hspace*{-1.1cm}\includegraphics[width=.95\textwidth,trim={0.3cm 0 0.3cm 0},clip]{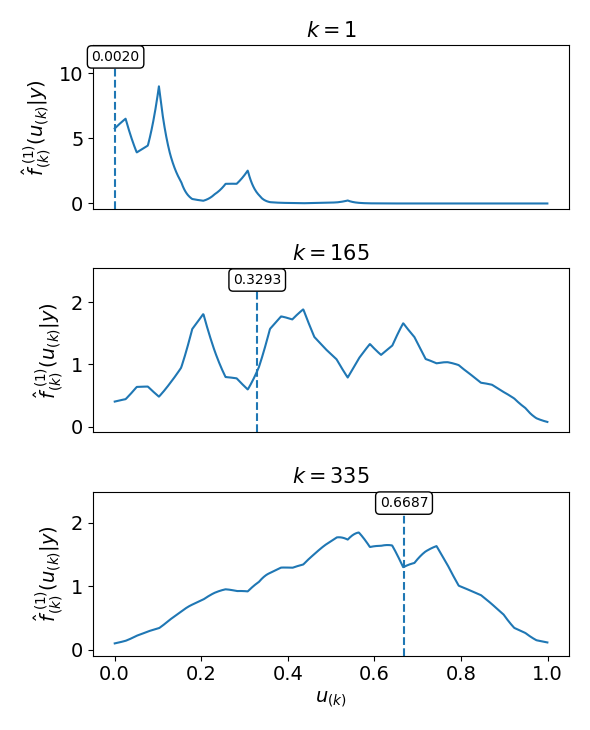}
		\end{minipage}
	\end{minipage}
	\caption{Graphon estimation based on linear B-splines and $\hat{\bu}^{(0)}$ for Graphon~1 (top left) from Table~\ref{tab:graFun}. The graphon estimate $\hat{w}^{(1)}(,)$ is depicted in top right. The plot at bottom left illustrates the estimated $\hat{u}_i^{(0)}$ versus the true simulated $u_i$. The three lower right plots show the approximated posterior distribution of $U_{(k)}$ (based on the MCMC samples and with respect to the given graphon estimate) for some selected indices. The dashed vertical lines (see also numbers in the boxes) represent the estimates $\hat{u}_{(k)}^{(0)}$.}
	\label{fig:sim1GraEst1}
\end{figure}

For evaluating our method we at first consider networks generated from a known ground truth. More precisely, for each of the two graphons from Table~\ref{tab:graFun} we simulate networks with dimension $N=500$ using the data generating process (\ref{eq:mod1}). 
\begin{table}
	\begin{center}
		\begin{tabular}{cc}
			\toprule
			ID \qquad & Graphon \\ \midrule
			1 \qquad & $\begin{aligned} w_1 (u,v) = 0.8 \left( 1-u \right) \left( 1-v \right) + 0.85 (u \cdot v) \end{aligned}$ \\[0.3cm]
			2 \qquad & $\begin{aligned} w_2 (u,v) = & \; 0.5 \cdot \{F_{\mathcal{N}(0,0.25)}( \Phi^{-1}(u)) \cdot F_{\mathcal{N}(0,0.25)}( \Phi^{-1}(v)) \\
			& + [1 - F_{\mathcal{N}(0,0.25)}( \Phi^{-1}(u)) ] \cdot [1 - F_{\mathcal{N}(0,0.25)}( \Phi^{-1}(v)) ]\} \end{aligned}$ \\ \bottomrule
		\end{tabular}
	\end{center}
	\caption{Exemplary graphons considered for simulations. For the second graphon specification $\Phi()$ and $F_{\mathcal{N}(0,0.25)}$ denote the cdf's of the normal distributions with parametrization $(\mu, \sigma^2) = (0,1)$ and $(0,0.25)$, respectively.}
	\label{tab:graFun}
\end{table}
The first graphon has a canonical representation, while the second is more general and does not fulfill that $g()$ in (\ref{eq:g}) is strictly increasing. We start with Graphon~1 and demonstrate the benefits of iteratively applying the E- and M-step instead of ordering the nodes just with respect to their degree, i.e. based on $\hat{\psi}()$, and applying the M-step only once. We call the latter a ``one-step'' estimator. This ``one-step'' estimate $\hat{w}^{(1)}(,)$ is shown in the top right panel in Figure~\ref{fig:sim1GraEst1}. 
The fit is apparently not convincing, which we want to make more evident by looking at the ranking information. First, we plot $\hat{u}^{(0)}_i$ as defined in (\ref{eq:uinit}) against the true values $u_i$. There is no concordance visible, which exhibits the problem with a simple ranking of the nodes based on the degree. 
We run an MCMC sample according to Section~\ref{subsec:BayApp} to obtain information about the ranking, as discussed in Section~\ref{subsec:infonrank}. The MCMC sample allows to estimate the posterior distribution of $U_{(k)}$ as proposed in (\ref{eq:postDist}), which for three exemplary values of $k$ is plotted in the bottom right plots of Figure~\ref{fig:sim1GraEst1}. The corresponding initial estimates $\hat{u}_{(k)}^{(0)}$ are included as vertical dashed lines and it can be seen that they are not well represented by the posterior distributions. We can conclude that this network, which has been generated from $w_1(,)$, is not suitable to sort by degree for achieving an appropriate estimate of the true node ordering with respect to the data generating model. 
Hence, also the ``one-step'' estimation approach results in a very poor fit of the underlying true graphon. 
We therefore iterate between the E- and the M-step. 

\begin{figure}
	\centering
	\begin{minipage}[t]{1\textwidth}
		\centering
		\begin{minipage}[t]{0.48\textwidth}
			\centering
			\includegraphics[width=.99\textwidth]{2m_real_graphon}
		\end{minipage}
		\hspace*{0.1cm}
		\begin{minipage}[t]{0.48\textwidth}
			\centering
			\includegraphics[width=.99\textwidth]{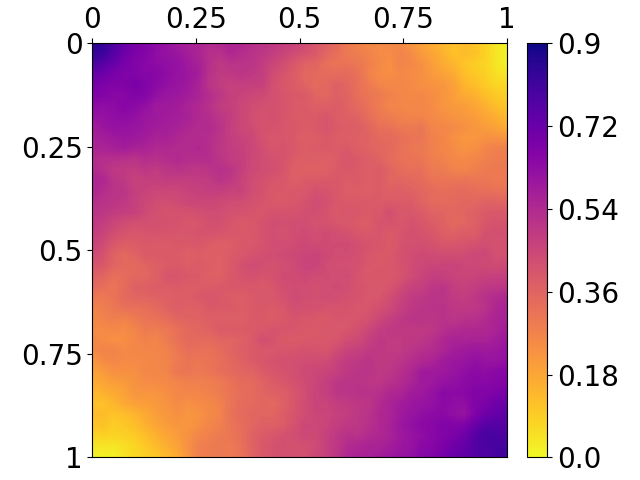}
		\end{minipage}
		\vspace*{0.3cm}
		
		\begin{minipage}[c]{0.49\textwidth}
			\centering
			\hspace*{-1.1cm}\includegraphics[width=.87\textwidth, height=.75\textwidth]{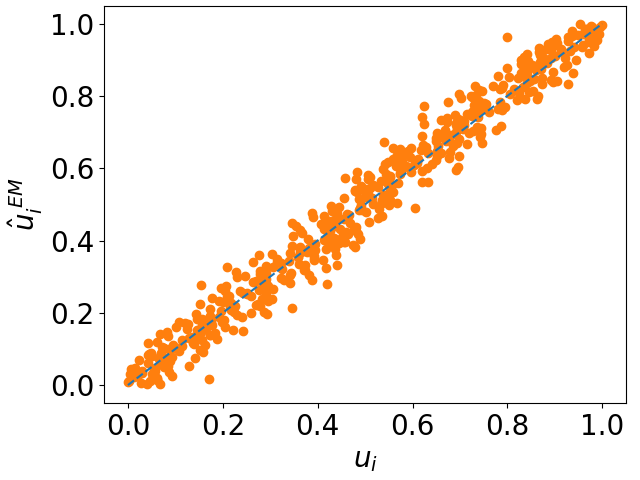}
		\end{minipage}
		\begin{minipage}[c]{0.49\textwidth}
			\centering
			\hspace*{-1.1cm}\includegraphics[width=.95\textwidth,trim={0.3cm 0 0.3cm 0},clip]{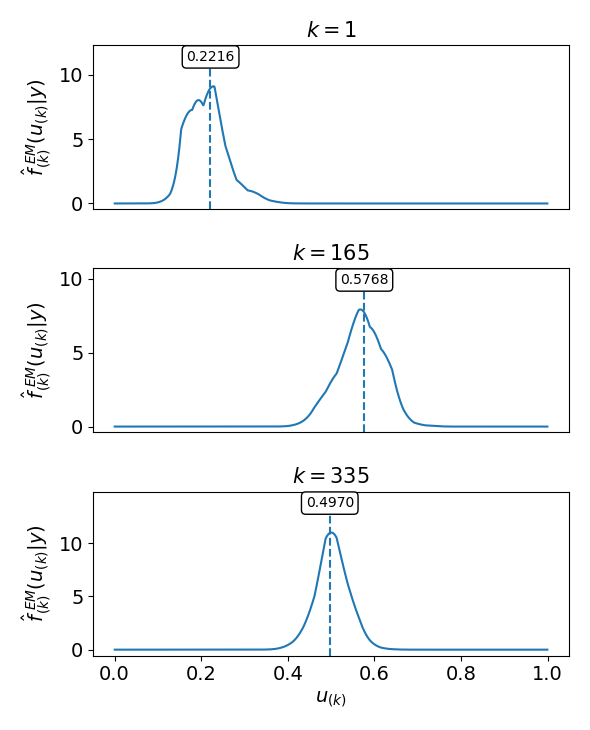}
		\end{minipage}
	\end{minipage}
	\caption{Graphon estimation for Graphon~1 (top left) from Table~\ref{tab:graFun}, based on the EM algorithm using $\hat{\bu}^{(0)}$ from (\ref{eq:uinit}) as initialization and excluding the canonical restriction from (\ref{eq:canonical}). The graphon estimate $\hat{w}^{\text{\textit{EM}}}(,)$ is depicted in the top right. 
		The plot at bottom left illustrates the comparison between the estimated $\hat{u}_i^{\text{\textit{EM}}}$ and the true simulated $u_i$. The three lower right plots show for some selected indices the approximated posterior distribution of $U_{(k)}$ with respect to the graphon estimate in the top right panel. The dashed vertical lines (see also numbers in the boxes) represent the estimates $\hat{u}_{(k)}^{\text{\textit{EM}}}$.}
	\label{fig:sim2GraEst2}
\end{figure}

The converged EM estimate is shown in the top right plot in Figure~\ref{fig:sim2GraEst2}. 
Regarding the final EM based estimates $\hat{u}_i^{\text{\textit{EM}}}$, the comparison with the true values $u_i$ reveals a convincing concordance. We also consider again the posterior distribution of $U_{(k)}$ for the same indices as above (bottom right plots), which indicates a plausible positioning. 
As a conclusion, the proposed EM algorithm provides appropriate results even if the initial ordering of the nodes by their degree is not adequate. Note that the canonical restriction from (\ref{eq:canonical}) has not been included here although $w_1(,)$ has canonical representation, since the shown unrestricted estimate turned out to be similarly good.

\begin{figure}
	\centering
	\includegraphics[width=.65\textwidth,trim={0 -0.3cm 0 1cm},clip]{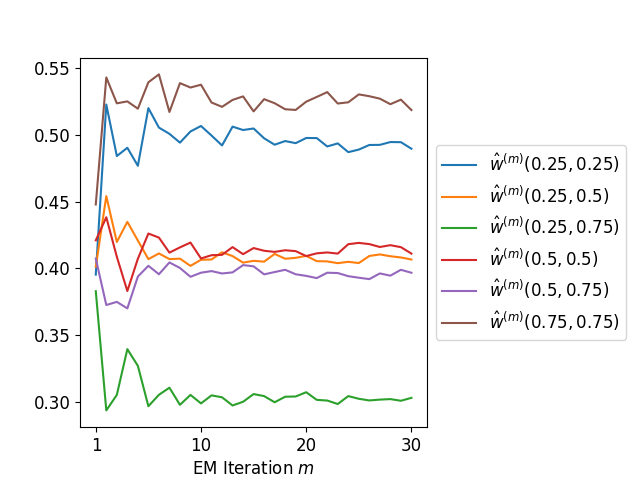}
	\caption{Trajectory of $\hat{w}^{(m)} (u,v)$ for Graphon~1 from Table~\ref{tab:graFun} at exemplary specific positions for the proceeding EM iterations $m=1,\ldots,30$.}
	\label{fig:sim1Traj}
\end{figure}

Finally, we explore the convergence behavior of the algorithm, which is illustrated in Figure~\ref{fig:sim1Traj} for some values of the graphon at exemplary specific positions. We see that after about 8 to 12 steps a reasonable convergence occurs. 

\begin{figure}
	\centering
	\begin{minipage}[t]{1\textwidth}
		\centering
		\begin{minipage}[t]{0.48\textwidth}
			\centering
			\includegraphics[width=.99\textwidth]{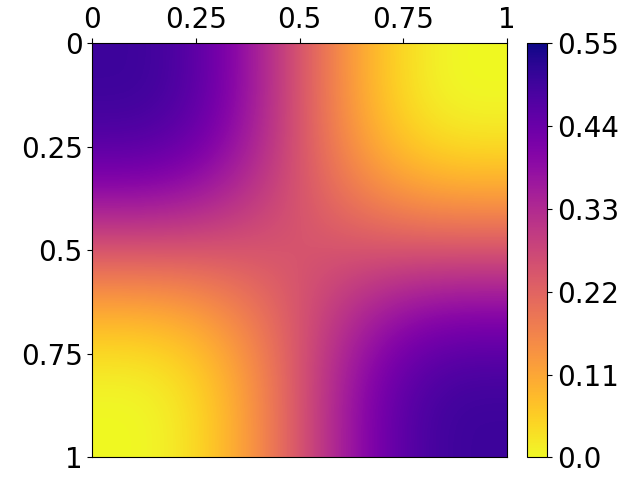}
		\end{minipage}
		\hspace*{0.1cm}
		\begin{minipage}[t]{0.48\textwidth}
			\centering
			\includegraphics[width=.99\textwidth]{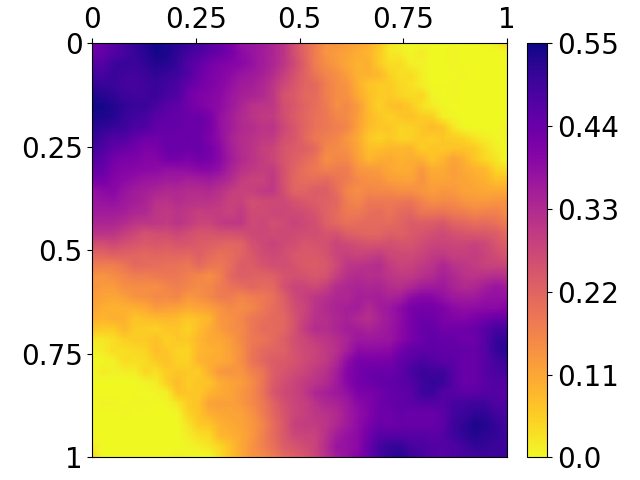}
		\end{minipage}
		\vspace*{0.3cm}
		
		\begin{minipage}[c]{0.49\textwidth}
			\centering
			\hspace*{-1.1cm}\includegraphics[width=.87\textwidth, height=.75\textwidth]{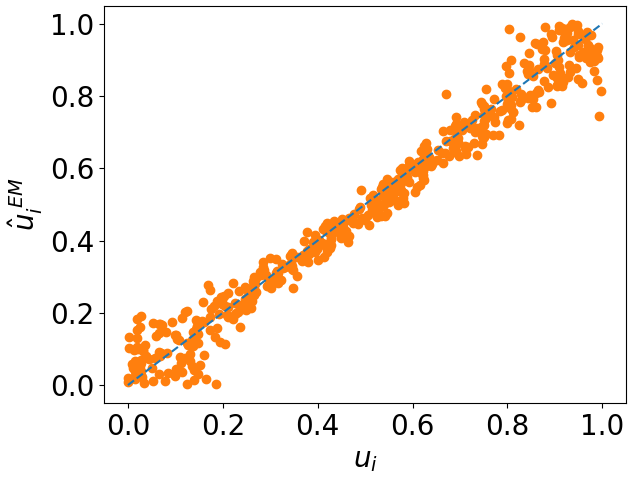}
		\end{minipage}
		\begin{minipage}[c]{0.49\textwidth}
			\centering
			\hspace*{-1.1cm}\includegraphics[width=.95\textwidth,trim={0.3cm 0 0.3cm 0},clip]{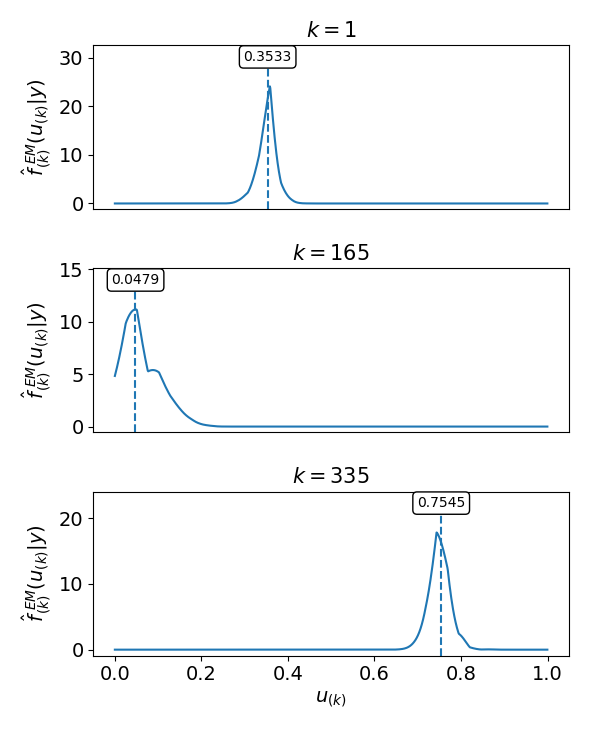}
		\end{minipage}
	\end{minipage}
	\caption{Graphon estimation for the non-canonical Graphon~2 (top left) from Table~\ref{tab:graFun}, based on the EM algorithm using $\hat{\bu}^{(0)}$ from (\ref{eq:uinit}) as initialization and excluding the canonical restriction from (\ref{eq:canonical}). 
		The graphon estimate $\hat{w}^{\text{\textit{EM}}}(,)$ is depicted in the top right. The plot at bottom left illustrates the estimated $\hat{u}_i^{\text{\textit{EM}}}$ versus the true simulated $u_i$. The three lower right plots show for some selected indices the approximated posterior distribution of $U_{(k)}$ with respect to the graphon estimate in the top right panel. The dashed vertical lines (see also numbers in the boxes) represent the estimates $\hat{u}_{(k)}^{\text{\textit{EM}}}$.}
	\label{fig:sim3GraEst2}
\end{figure}

\begin{figure}
	\centering
	\hspace{-1.0cm}
	\begin{minipage}[t]{0.32\textwidth}
		\centering
		\includegraphics[width=1.1\textwidth,trim={1.4cm 1cm 1.0cm -0.5cm},clip]{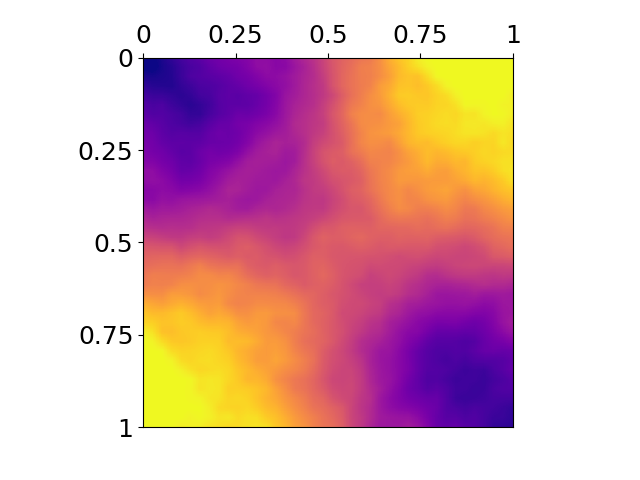}
		\vspace*{-0.3cm}
		
		\hspace*{0.55cm}{\scriptsize $\begin{gathered} 
			l_{\bP} (\btheta, \lambda) = -123,291.08 \\ AIC_c(\lambda) = 247,369.06
			\end{gathered}$}
	\end{minipage}
	\hspace{-0.3cm}
	\begin{minipage}[t]{0.32\textwidth}
		\centering
		\includegraphics[width=1.1\textwidth,trim={1.4cm 1cm 1.0cm -0.5cm},clip]{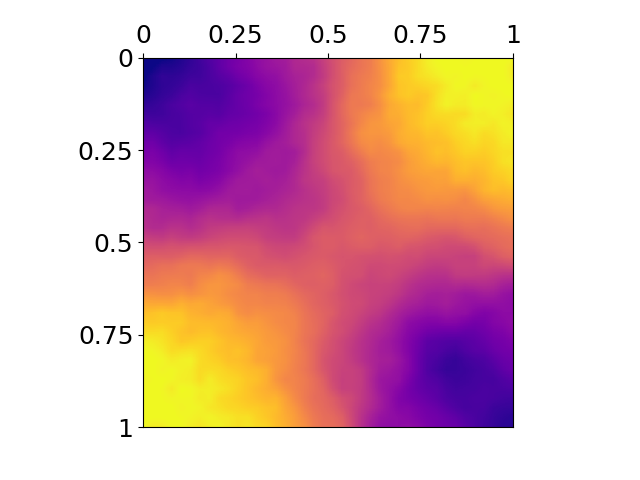}
		\vspace*{-0.3cm}
		
		\hspace*{0.55cm}{\scriptsize $\begin{gathered} 
			l_{\bP} (\btheta, \lambda) = -123,447.84 \\ AIC_c(\lambda) = 247,582.89
			\end{gathered}$}
	\end{minipage}
	\hspace{-0.1cm}
	\begin{minipage}[t]{0.32\textwidth}
		\centering
		\includegraphics[width=1.1\textwidth,trim={1.4cm 1cm 1.0cm -0.5cm},clip]{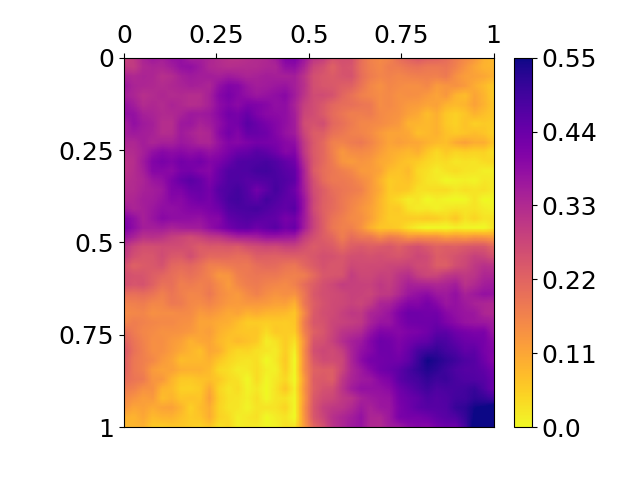}
		\vspace*{-0.3cm}
		
		\hspace*{0.2cm}{\scriptsize $\begin{gathered} 
			l_{\bP} (\btheta, \lambda) = -124,530.64 \\ AIC_c(\lambda) = 250,136.38
			\end{gathered}$}
	\end{minipage}
	\vfill
	\hspace{-1.0cm}
	\begin{minipage}[t]{0.32\textwidth}
		\centering
		\includegraphics[width=1.1\textwidth,trim={1.4cm 1cm 1.0cm -0.5cm},clip]{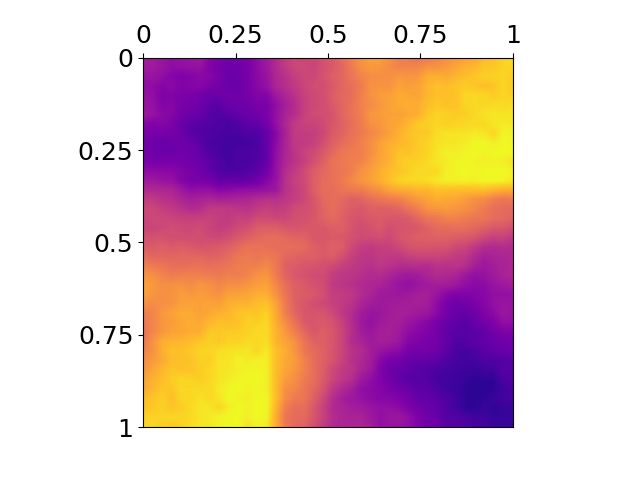}
		\vspace*{-0.3cm}
		
		\hspace*{0.55cm}{\scriptsize $\begin{gathered} 
			l_{\bP} (\btheta, \lambda) = -123,830.82 \\ AIC_c(\lambda) = 248,469.47
			\end{gathered}$}
	\end{minipage}
	\hspace{-0.3cm}
	\begin{minipage}[t]{0.32\textwidth}
		\centering
		\includegraphics[width=1.1\textwidth,trim={1.4cm 1cm 1.0cm -0.5cm},clip]{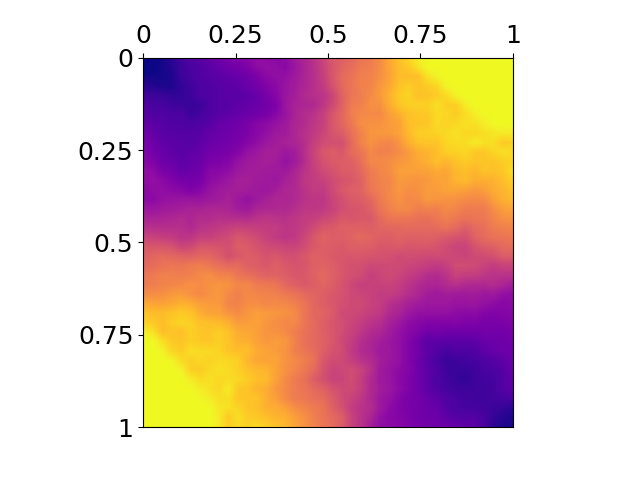}
		\vspace*{-0.3cm}
		
		\hspace*{0.55cm}{\scriptsize $\begin{gathered} 
			l_{\bP} (\btheta, \lambda) = -123,247.88 \\ AIC_c(\lambda) = 247,344.62
			\end{gathered}$}
	\end{minipage}
	\hspace{-0.1cm}
	\begin{minipage}[t]{0.32\textwidth}
		\centering
		\includegraphics[width=1.1\textwidth,trim={1.4cm 1cm 1.0cm -0.5cm},clip]{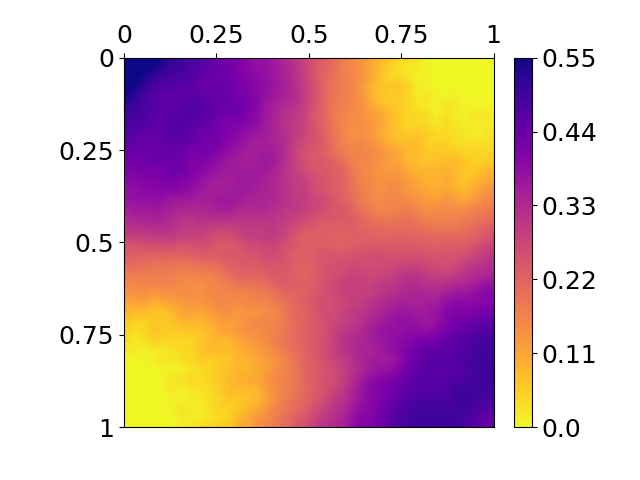}
		\vspace*{-0.3cm}
		
		\hspace*{0.2cm}{\scriptsize $\begin{gathered} 
			l_{\bP} (\btheta, \lambda) = -123,431.02 \\ AIC_c(\lambda) = 247,556.45
			\end{gathered}$}
	\end{minipage}
	\caption{Graphon estimates $\hat{w}^{\text{\textit{EM}}}(,)$ for the non-canonical Graphon~2 from Table~\ref{tab:graFun}, based on the EM algorithm using uninformative random initialization. For all repetitions the same simulated network has been used but with independently differing initialization. Beneath are the penalized likelihood and the corrected AIC given accordingly.}
	\label{fig:sim2GraEst11}
\end{figure}

For a further evaluation, we now have a look at Graphon~2, which does not provide a canonical representation and thus is completely intractable for simple degree based estimation methods. This simulation seeks to demonstrate that our proposed method can also handle more general graphons. The marginal function $g()$ as defined in (\ref{eq:g}) is constant at $0.25$, meaning that the degree is fully uninformative with regard to the node ordering. Nonetheless, we here remain with the degree based initialization as proposed above to show that the EM algorithm is capable of appropriately estimating the underlying structure even under bad initialization. 
We exclude the side constraint from (\ref{eq:canonical}), since we now explicitly want to enable flexible marginal functions. The resulting graphon estimate for a simulated network of dimension $N=500$ is depicted in the top right plot in Figure~\ref{fig:sim3GraEst2}. 
The structure of the true graphon (top left) is fully captured, which is accompanied by a convincing concordance between the estimated $\hat{u}^{\text{\textit{EM}}}_i$ and the true $u_i$. In accordance with that, the estimates $\hat{u}^{\text{\textit{EM}}}_{(k)}$ are well covered by the corresponding ranking densities, which is illustrated for three selected indices at the stacked plots at the bottom right. To further showcase the performance of our method we repeat the estimation procedure several times based on the same simulated network but with differing random initialization. More precisely, instead of making use of the degree based ordering from (\ref{eq:uinit}), we set $\hat{\bu}^{(0)}$ as a random permutation of $(i/(N+1):\; \allowbreak i =1 ,\ldots, N)$. This also exemplary characterizes the appearance of the EM algorithm being trapped at local minima of the AIC. Figure~\ref{fig:sim2GraEst11} illustrates the estimation results for six such repetitions. 
Although we start for each run with a completely uninformative random initialization, in most of the final estimates the structure of the original graphon can be fully recognized instantly. Yet, the estimates in the top right and the bottom left panel reveal a somehow differing structure. Having a more precise look, they exhibit the appearance of segment swaps and reversals with respect to the original graphon representation. For instance, at the graphon estimate in the bottom left panel the upper and the lower part of the domain $[0,1]$ have been swapped and, in addition, the lower part has been reversed. However, as mentioned before, applying permutations to the graphon function has no effect on the model specification itself. Looking at the resulting AIC values beneath the graphon estimates, we nevertheless see that the smallest values result for smooth graphon estimates, while the occurrence of a ``jump'' due to only piecewise correct reordering of the nodes lead to increased AIC values. 
But still, the structure of the true graphon model is always well captured, albeit possibly in a different form of representation. Summarizing over all the estimation results, our algorithm yields very good estimates for Graphon~2, although an uninformative initial node ordering has been applied. 

The example illustrates that a random reordering of the nodes for the initial E step can be applied in combination with the AIC in order to obtain an optimal smooth graphon estimate.

\section{Real-World Data Examples}
\label{sec:realWorld}
We complete the paper by analyzing real-world network data. To do so, we consider networks from three different domains, namely from sociology, political science, and neuro\-science. For the EM based estimation we use degree ordering as initialization where the structure seems appropriate. Otherwise we apply random initialization and select the best outcome over several repetitions.

\subsection{Facebook Network}
\label{sub:facExp}
\begin{figure}[p]
	\centering
	\includegraphics[width = 0.9\textwidth]{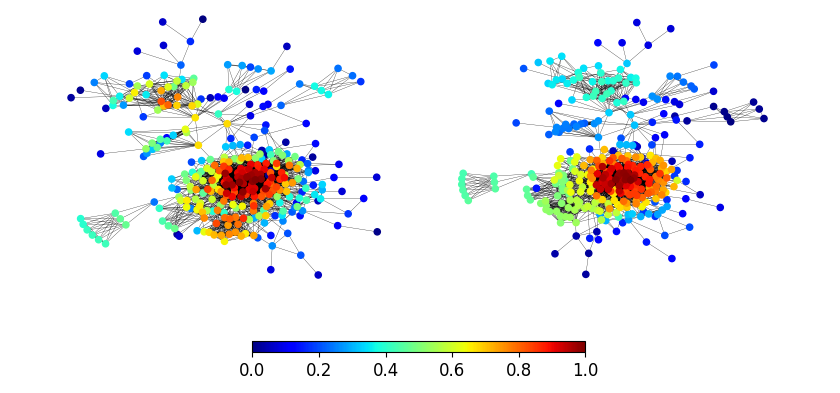}
	\caption{Facebook ego network with node coloring referring to $\hat{u}_i^{(0)}$ (left) and $\hat{u}_i^{\text{\textit{EM}}}$ (right).}
	\label{fig:fbNetwork}
\end{figure}

\begin{figure}
	\centering
	\begin{minipage}[c]{0.52\textwidth}
		\centering
		\hspace*{-1.1cm}\includegraphics[width=.99\textwidth,trim={1.6cm 0 0.3cm 0},clip]{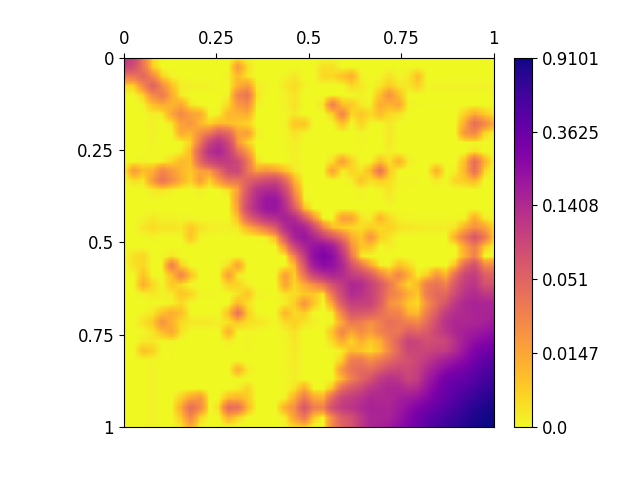}
	\end{minipage}
	\begin{minipage}[c]{0.47\textwidth}
		\centering
		\hspace*{-1.1cm}\includegraphics[width=.95\textwidth,trim={0.3cm 0 0.3cm 0},clip]{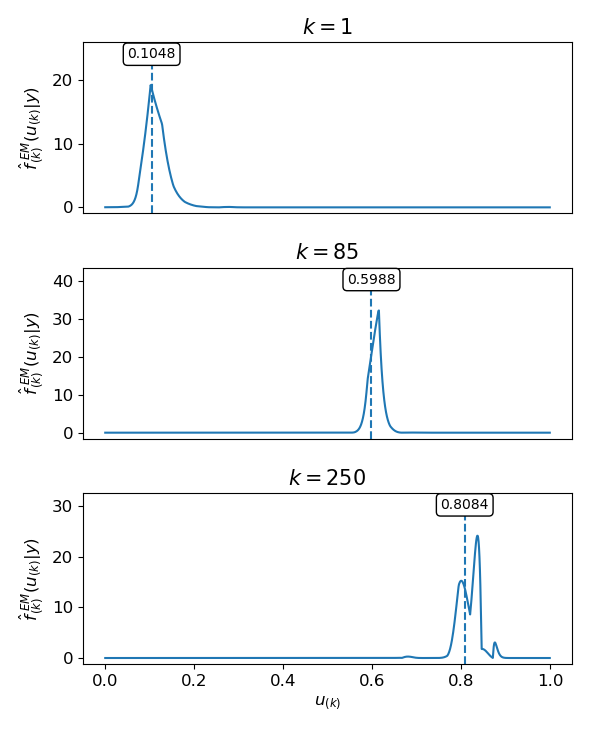}
	\end{minipage}
	\caption{Graphon estimation for the Facebook ego network. The graphon estimate $\hat{w}^{\text{\textit{EM}}}(,)$ (in log scale) is depicted on the left. The three right plots show the approximated posterior distribution of $U_{(k)}$ (with respect to the given graphon estimate) for some selected indices. The dashed vertical lines (see also numbers in the boxes) represent the estimates $\hat{u}_{(k)}^{\text{\textit{EM}}}$.}
	\label{fig:fbPostDist2}
\end{figure}

A very common application and one of the roots of network sciences are friendship networks. We here consider a Facebook ego network which has been collected by \cite{McAuley:12} and is available on the Stanford Large Network Dataset Collection \citep{snapnets}. This ego network with $333$ actors is plotted in Figure~\ref{fig:fbNetwork} with two different colorings. 
In the left panel the illustrated ordering is based on the degree, while on the right it refers to the final EM result. Apparently, the ordering given by the final EM result seems much more appropriate with respect to the network structure. Thus, the iterative approach improves the initial degree ordering markedly. Moreover, the inherent structure of the network on the right in Figure~\ref{fig:fbNetwork} can to some extent be recognized instantly in the corresponding final graphon estimate $\hat{w}^{\text{\textit{EM}}}(,)$, which is shown on the left in Figure~\ref{fig:fbPostDist2}. 
For example, the bundle of nodes in the center of the lower network part with roughly $\hat{u}_i^{\text{\textit{EM}}} \in [0.65,1]$ and the connectivity pattern among themselves is captured in the graphon estimate in the intense section on the bottom right. 
In accordance with this, also the estimates $\hat{u}_{(k)}^{\text{\textit{EM}}}$ for some selected indices are adequately represented by the corresponding posterior distributions of $U_{(k)}$, which underlines the appropriateness of the graphon estimate $\hat w^{\text{\textit{EM}}} (,)$.

\subsection{Military Alliance Network}
\begin{figure}
	\centering
	\begin{minipage}[t]{1\textwidth}
		\centering
		\hspace*{0.8cm}\includegraphics[width=.93\textwidth]{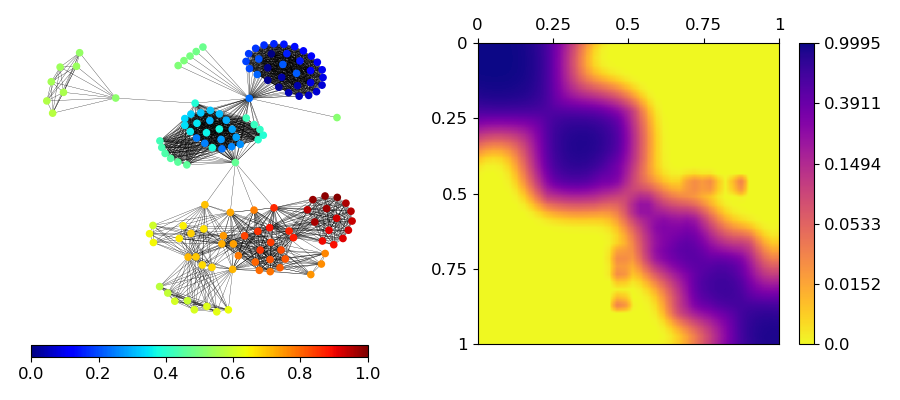}
	\end{minipage}
	\vspace*{-0.1cm}
	
	\begin{minipage}[b]{1\textwidth}
		\centering
		\includegraphics[width=.99\textwidth,trim={0 3cm 0 3cm},clip]{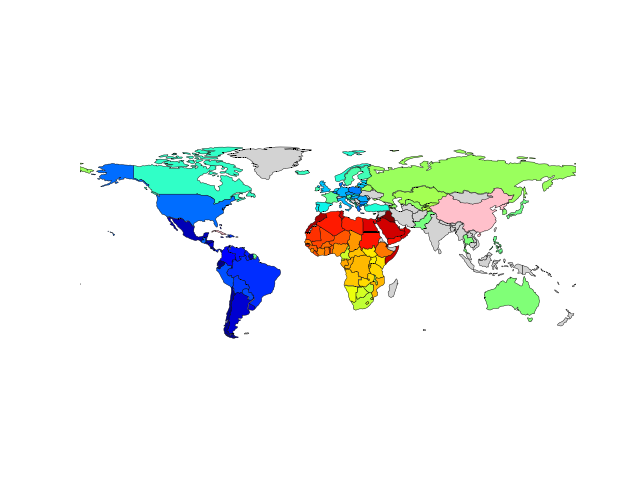}
	\end{minipage}
	\caption{Graphon estimation for the military alliance network (top left, with coloring referring to $\hat{u}_i^{\text{\textit{EM}}} \in [0,1]$). The graphon estimate $\hat{w}^{\text{\textit{EM}}}(,)$ (in log scale) is depicted in the top right. The lower plot shows the world map, where the colors also refer to $\hat{u}_i^{\text{\textit{EM}}}$. As an isolated group, China, Cuba, and North Korea (colored in pink) have not been included in the estimation routine. Countries which have no strong agreement with any other country and hence do not appear in the data set are colored in gray.}
	\label{fig:alliance}
\end{figure}

As second real-world network example, we consider strong military alliances among the world's nations. For that purpose, we use data from the Alliance Treaty Obligations and Provisions project \citep{Leeds:02}, which provide information about all kind of military alliance agreements over an extensive period. For a suitably reduced network we here define the presence of a strong military alliance when states have entered into an offensive or defensive pact, meaning when they have signed a treaty which forces the one country to intervene by military active support if the other country comes into a conflict with offensive or defensive actions, respectively. Furthermore, we truncate the data to agreements that were in force in 2016 as the most recent available year. The best final estimation results of the EM approach over several repetitions with random initial node ordering are illustrated in Figure~\ref{fig:alliance}. 
The graphon estimate in the top right panel exhibits a very pronounced assortative structure, meaning that predominantly pairs of nodes are linked whose latent quantities are close. The final node ordering, which is visualized by the coloring in the top left network, appears reasonable with respect to the network structure. Transferring this coloring to the world map reveals a strong conformity between the geographic closeness and the closeness of the latent quantities. Together with the assortative structure, this means that countries which are geographically close form similar military alliances and further are more likely to be allied with each other.

\subsection{Human Brain functional Coactivation Network}
\begin{figure}[t]
	\centering
	\begin{minipage}[t]{1\textwidth}
		\centering
		\hspace*{0.8cm}\includegraphics[width=.93\textwidth]{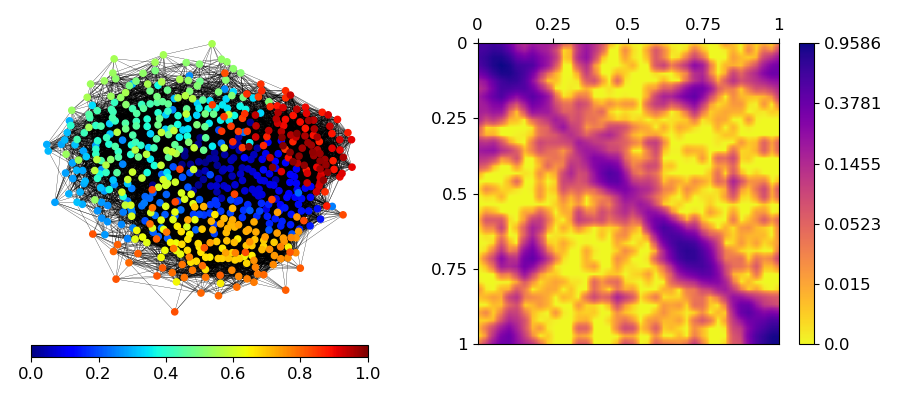}
	\end{minipage}
	\vspace*{-0.1cm}
	
	\begin{minipage}[t]{0.32\textwidth}
		\centering
		\includegraphics[width=.98\textwidth,trim={1.5cm 0.5cm 1.5cm 0.5cm},clip]{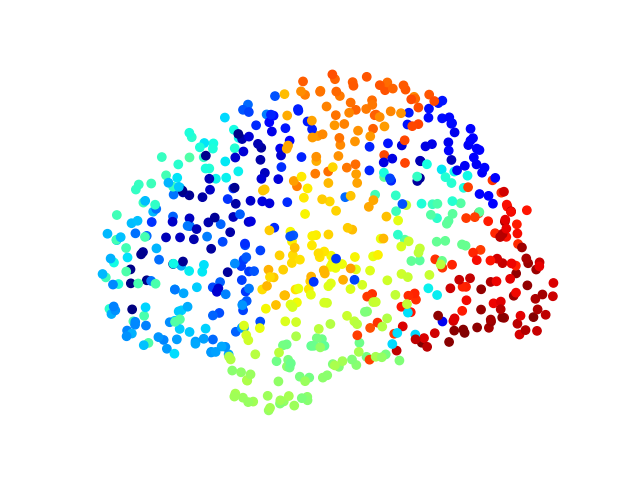}
	\end{minipage}
	\begin{minipage}[t]{0.32\textwidth}
		\centering
		\includegraphics[width=.98\textwidth,trim={1.5cm 0.5cm 1.5cm 0.5cm},clip]{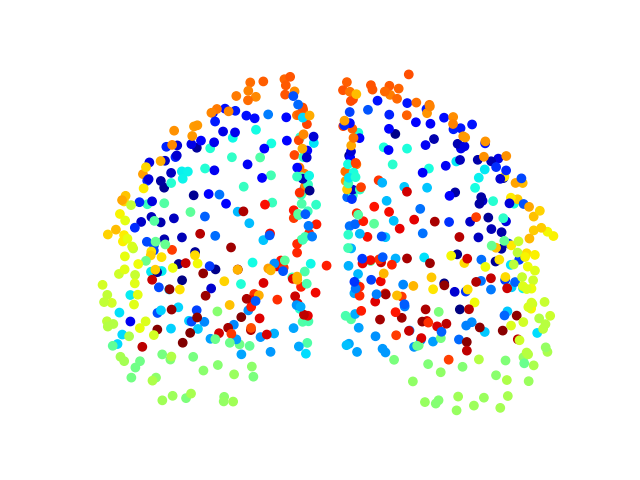}
	\end{minipage}
	\begin{minipage}[t]{0.32\textwidth}
		\centering
		\includegraphics[width=.98\textwidth,trim={1.5cm 0.5cm 1.5cm 0.5cm},clip]{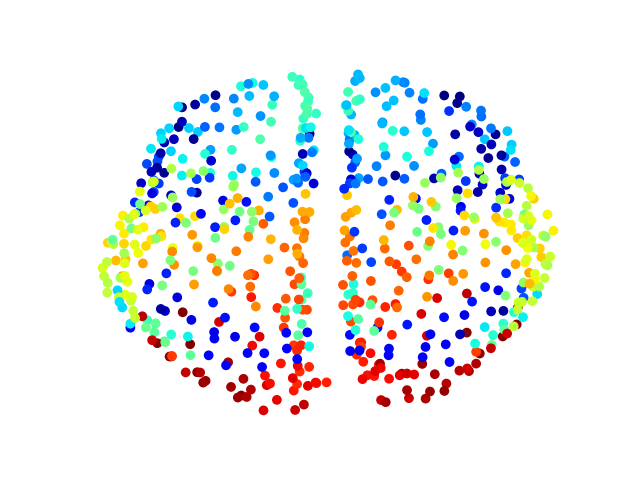}
	\end{minipage}
	\caption{Graphon estimation for the human brain functional coactivation network (top left, with coloring referring to $\hat{u}_i^{\text{\textit{EM}}} \in [0,1]$). The graphon estimate $\hat{w}^{\text{\textit{EM}}}(,)$ (in log scale) is depicted in the top right. The lower three plots show the local positions of the human brain regions in anatomical space in side view (left), front view (middle), and top view (right), where the colors also refer to $\hat{u}_i^{\text{\textit{EM}}}$.}
	\label{fig:brain}
\end{figure}

To conclude the real-world examples, we consider a human brain functional network which has been constructed from a meta-analysis by \cite{crossley:13} and which is available in the Brain Connectivity Toolbox (\citeauthor{brain:10}, \citeyear{brain:10}, see also \citeauthor{rubinov:10}, \citeyear{rubinov:10} for detailed description). Their weighted network matrix represents the ``estimated [\ldots] similarity (Jaccard index) of the activation patterns across experimental tasks between each pair of $638$ brain regions'' \citep{crossley:13}. To obtain a binary adjacency matrix out of it, we simply apply a threshold of above zero, meaning that we include a link between each pair of brain regions which is at least during one task activated simultaneously. The resulting network with a density of approximately $9.2\%$ is depicted in the top left panel in Figure~\ref{fig:brain}, where the coloring refers to $\hat{u}_i^{\text{\textit{EM}}}$. 
For the evaluation, we again consider the best outcome of the EM algorithm over several repetitions with random initialization. The graphon estimate in the top right plot also reveals an assortative structure but, in addition, exhibits a conspicuous pattern of functional coactivations of some segments, which are separated with respect to the latent dimension. Regarding the spatial positions of the brain regions, the lower three plots show a strong relation between closeness in anatomical space and closeness of the $\hat{u}_i^{\text{\textit{EM}}}$. Nevertheless, there also seem to be areas which have a similar color pattern (indicating a similar coactivation pattern) but are spatially separated. Indeed, the interaction and coactivation of distant brain areas is a familiar phenomena in neuroscience and can also be seen in \cite{crossley:13}, who pursue a different modeling strategy. Overall, we can demonstrate that the EM approach for estimating smooth graphons provides additional insight into network structures.

\section{Discussion}
The paper proposes a novel estimation routine for graphon estimation which explicitly  takes the variability of ordering the nodes into account. The proposed semi-parametric estimation based on (linear) B-splines allows to incorporate relevant properties in the estimation and also uniqueness restrictions if required. The Bayesian approach relying on Gibbs sampling illuminates the uncertainty about the degree and its distribution. Both steps combined give an EM type algorithm which allows for flexible graphon estimation even in large networks. The approach outperforms available routines in two aspects. First, the B-spline estimate can guarantee a smooth and continuous outcome and also a unique representation of the graphon if required. Secondly, based on the Bayesian formulation and the EM algorithm one can assess the amount of uncertainty for ordering the nodes based on their degree or based on any other strategy.

The proposed approach can also be used in other related models like stochastic block models (SBM), where we assume that nodes cluster and form within and between the clusters simple Erd\H{o}s-R\'{e}nyi models. Apparently, SBMs do not have a smooth underlying graphon structure so that one needs to adjust and further develop the results in this paper. 

Overall, graphon estimation also provides an interesting tool for network visualizations, as demonstrated in the examples. In this respect, it is more than a modeling exercise but may also serve as tool for explanatory network data analysis.

\if0\blind
{
	\section*{Acknowledgments}
	
	The project was partially supported by the European Cooperation in Science and Tech\-nology [COST Action CA15109 (COSTNET)]. This research did not receive any specific grant from funding agencies in the public, commercial, or not-for-profit sectors. 	
	Declarations of interest: none.
	
} \fi

\bibliographystyle{chicago}

\bibliography{bibliography}

\end{document}